\documentclass[review]{elsarticle}

\usepackage{lineno,hyperref}
\modulolinenumbers[5]

\journal{arXiv}

\begin{document}

\begin{frontmatter}

\title{A higher-dimensional Lagrangian representation of bi-cylindrical dynamical geometry for the mass-hierarchy problem of charged leptons}

\author{Vo Van Thuan}
\address{Duy Tan University (DTU)\\
3 Quang Trung street, Hai Chau district, Danang, Vietnam\\
Vietnam Atomic Energy Institute (VINATOM)\\
59 Ly Thuong Kiet street, Hoan Kiem district, Hanoi, Vietnam}

\begin{abstract}
According to our microscopic cosmological model, masses of charged leptons can be calculated by curvatures of hyper-spherical surfaces embedded in a 3D time-like subspace. In this study, a higher-dimensional Lagrangian representation is applied to the bi-cylindrical dynamical geometry. A four-component scalar field is assumed to serve as the gauge transformers in the linearly factorized 4D space-time for inducing lepton mass. The consistency of the general relativity approach with quantum field theories is discussed. 

\end{abstract}


\end{frontmatter}


\section{Introduction}
Mass hierarchy is an important problem of particle physics. 
In the Standard model (SM),  elementary particle masses are induced through the interaction of genetic particles  with the Higgs field characterized by a global nonzero averaged vacuum expectation.  However, each kind of particle has its own Yukawa coupling of unknown origin.
Among fermion elementary particles, the masses of quarks are calculated not only as pole-massive quantities \cite{Big1}, but also as running masses by effective field theories dependent on accepted realistic energy cutoff thresholds up to the electro-weak scale \cite{{Elk1},{XiZhaZh1}}. Except Yukawa couplings, they need an extended parametrization by mixing CKM matrices \cite{{Cab1},{KoMa1}}, being in consistency with QCD, within the SM (for a review, see, e.g. \cite{Hoc1}). Based on the idea of neutrino oscillation suggested by Pontecorvo \cite{Pon1} another mixing scheme has been applied by parametrization of MNSP matrices \cite{{MNS1},{Pon2}} which leads to larger mixing angles. Even though the oscillation experiments confirmed non-zero masses of neutrinos, their absolute masses remain unknown and are considered as a beyond SM problem. In opposition, the absolute masses of charged leptons are measured with very high accuracy \cite {Patr1} and all experimental facts show that any flavor violation which would lead to their mass mixing, should be suppressed in SM. The latter, if exists, would open new physics of beyond SM at muon factories (see, e.g., \cite{Oot1} for a review). \\
In general, the origins of the mass hierarchy of charged leptons, neutrinos and quarks would be assumed independent, due to a significant mismatch between different sectors. Meanwhile, there are other trends to consider that the mass hierarchy of all elementary fermions would have partially common origins, which lead to some of their relationships. In phenomenological approaches within an extension of the SM, there are suggestions, for example, that they would be governed by some common higher symmetry which is broken spontaneously to get different orders of mass ratios in different elementary particle sectors. In particular, the idea of "flavor democracy" \cite {HaHa1} would give a correlation between original masses  of the up-type and down-type quarks \cite {{BrSi1},{FriXi1},{FiHu1},{XiYaZh1}}.   \\
A common origin would be implied even through an assumption of quark-to-lepton correspondences \cite {{XiYaZh1},{Koi2},{Koi3},{HoSa1},{FriXi2}}, in which the mass ratios of charged leptons would be determined by applying an assumed weakly mixing angles in a correlation with the quark sector or neutrino one.
For example, an analogue of the empirical formulas of quark mixing is applied to the mixing of both neutrino and charged lepton sectors \cite {FriXi2}. 
While all charged fermion elementary particles definitely have a normal ordering of mass hierarchy, it is unknown, whether the ordering of neutrino masses is normal or inverse.
The fact that two mixing angles $\theta_{23}$ and $\theta_{12}$ are observed as very large while the third angle $\theta_{13}$ is relatively small leads to an assumption of a weakness of the neutrino mass hierarchy, which would be expected from a moderate extension of SM with flavor symmetries \cite{{FriXi3},{FriXi4},{Ishi1},{Alta1}}. In particular, a so-called tri-bi-maximal mixing \cite{Har1} has been improved by formulation of a non-zero $\theta_{13}$ and CP violating parameters \cite{{Xin1},{Ky1}}. However, it remains unseen as to why the masses of three neutrino generations too close to each other.\\
Returning back to the most experimentally studied charged lepton sector, being almost without flavor mixing, then their mass hierarchy is still a big puzzle. As above-mentioned, the origin of their Yukawa coupling constants is unknown and in the frame of SM most efforts to solve their mass hierarchy problem are phenomenological. In the past, Barut proposed a conjecture that a magnetic self-interaction of electron induces masses of muon, tauon and heavier leptons as specific electron quantum excited states \cite{Bar1}, which leads roughly to their mass hierarchy. However, no heavier charged leptons have been detected. Koide discovered that the masses of charged leptons are related to each other in a quite simple but still unexplained empirical formula \cite{Ko1}.
Generally, it is accepted that the problem of mass hierarchy of charged leptons seems to be solved beyond SM. Recently, managing to conserve the Koide formula, Sumino \cite{Sumi1} proposed to search for its origin by an extension to the higher $U(3)$x$SU(2)$ family gauge symmetry at very high energies which leads approximately to the SM as to an effective field theory (EFT).\\
Following the trend of beyond SM with more alternative solutions, higher dimensional general relativity would be introduced. For the modern Kaluza-Klein (K-K) models, extradimensions (EDs) are not compact, but would be extended to a macroscopic scale to introduce new physics. Arkani-Hamed, S. Dimopoulos and G. Dvali \cite{Dva1} pioneered for the brane model approach  with introduction of the space-like large EDs. The 5D AdS brane theory applied by Randall and Sundrum \cite {Ran1} to deal with the huge mass hierarchy between the Planck scale and electro-weak physics. In another approach of induced matter models, Wesson, Ponce de Leon, Mashhoon and Liu \cite{{We1},{Ma1}} proposed a space-time-matter theory (5D-STM) describing proper mass as a special time-like ED, which in a link with 4D space-time physics of elementary particles led to a qualitative interpretation of quantum mechanics  \cite{{We3},{We4}}. \\
Our recent study \cite{{Vo1},{Vo3}}, following the induced-matter approach, was based on time-space symmetry with two time-like EDs being made explicit as the proper time $t_0$ and the amplitude of the quantum wave function $\psi$. A duality was found between the quantum wave equation in 4D space-time and a relativistic geodesic description of curved higher dimensional time-space. In particular, the higher-dimensional geodesic solution showed some advantages for more quantitative interpretation of quantum mechanics than ones from the STM theory \cite{{We3},{We4}}. Based on this duality our strategy is to use the higher-dimensional geodesic dual sub-solution as a complementary, more geometrical and classical dynamics to extend our power in quantum mechanical interpretations of the wave-like dual sub-solution.
Moreover, in application of the higher-dimensional geodesic equations, the mass hierarchy problem of charged leptons would be solved by formulation of 3D time-like geometrical dynamical configurations being considered in a correlation with number three of lepton generations, in particular, tauon mass was predicted with an accuracy within $2.2\%$ \cite{Vo3}. This result seems to be moving in the right direction, but is not yet satisfactory as the existing experimental mass of tauon is known to have a much higher accuracy ($\leq 10^{-4}$). For  improving the predictability of the proposed microscopic cosmological model, some corrections to the highest curvatures are added from  higher-order approximations in \cite{Vo10}. \\
In the present study, a higher-dimensional Lagrangian representation is applied to the bi-cylindrical dynamical geometry, assuming linearization of the higher-dimensional geometry by a space-time factorization in inducing proper masses of charged leptons. The consistency of the general relativity approach with quantum field theories is discussed.   The article is organized as follows: in Section 2, the ED bi-cylindrical geometrical dynamics with time-space symmetry is introduced; in Section 3 this higher-dimensional geometry is treated under Lagrangian formalism to search for its consistency with quantum field theories; then in Section 4 the solution by the microscopic cosmological model is reviewed for the mass hierarchy problem of charged leptons in accordance with Ref.\cite{{Vo3},{Vo10}}; in Section 5 there some evidences for consistency of the proposed model with the basic concepts of QFT are subjects of analysis and the main achievements of the present model are discussed.
\section{Dual solutions of the higher-dimensional gravitational equation in time-space symmetry}
In accordance with Ref.\cite{Vo3}, there two time-like extra-dimensions added to 4D-spacetime. As a result, being linked to each other, the two orthonomal symmetrical time-like and space-like 3D-subspaces formulate an ideal 6D-flat $\{3T,3X\}=\{t_1,t_2,t_3{|}x_1,x_2,x_3\}$ symmetrical time-space:
\begin{equation}
  dS^2=dt_k^2-dx_l^2.
  \label{eq1}
  \end{equation}
Then, it is proposed that our physics is carried out strictly on its time-space symmetrical "lightcone":
  \begin{equation}
  dt_k^2=dx_l^2.
  \label{eq2}
  \end{equation}
where $k,l=1 \div 3$ are summation indices of coordinates in the flat time-space. The natural units ($\hbar=c=1$) are used commonly unless an explicit quantum dimension is introduced. 
Equation $(\ref{eq2})$ accompanies the balance of corresponding derivatives as: $\frac{\partial \psi_0 }{\partial t_k}=\frac{\partial \psi_0 }{\partial x_l}$,
 where $\psi_0(t_k,x_l)$ is a function of relative displacements. In particular, in the lightcone manifold $(\ref{eq2})$
 a harmonic correlation $\psi_0[dt_k,dx_l]$ between a space-like and a time-like differentials-displacements near a balancing position would be considered as:
 \begin{equation}
 \frac{\partial^2 \psi_0 }{\partial t_k^2}=\frac{\partial^2 \psi_0 }{\partial x_l^2}.
 \label{eq3}
 \end{equation}
Then $\psi_0(t_k,x_l)=\psi_0 e^{-i(\Omega_k t_k-k_l x_l)}$ with a constant amplitude resembles the harmonic plane wave function of a higher-dimensional massless
field, describing balanced extremely longitudinal fluctuations in time and space along their axes of linear translation. We assume that those isotropic and homogeneous plane waves serve as 
primitive dynamical sources of quantum fluctuations of space-time.  All chaos of displacements  $dx_l$ and $dt_k$ formulate square-averaged space-like and time-like potentials $V_X$ and $V_T$, respectively.\\
The symmetrical cylindrical geometry
$\{3T,3X\}\equiv \{\psi(t_0,t_k),\varphi(t_0,t_k),t_k \mid \\ \psi(x_n,x_l),\varphi(x_n,x_l),x_l\}$ is applied as the most simplified model for description of both motions of linear translation and rotation. In particular, it is convenient for description of fermions with only two opposite longitudinal projections of spin. 
In the 6D-lightcone, cylindrical variables $\{\psi,\varphi \}$ become explicit functions of linear coordinates $\{t_k,x_l\}$ and two 3D-local
affine parameters $x_n$ and $t_0$ which are introduced according to projection of the space-like conventional spin $\vec{s}$ and the time-like pseudo-spins $\vec{\tau}$, respectively. \\
The 6D geodesic is constrained by 3D-local cylindrical dynamics embedded in the 6D-flat vacuum $dS=0$ which leads to 6D bi-cylindrical geometry:
\begin{equation}
d\Sigma^2=ds^2-d\sigma^2=dt_i^2-dx_j^2,
\label{eq5}
\end{equation}
where $i,j=1\div 3$ are summation indices of cylindrical curved time-space. 
The 6D general relativity equation in vacuum reads:
\begin{equation}
R^m_q-\frac{1}{2}\delta^m_q R=0,
\label{eq7}
\end{equation}
which leads to an equivalent $\{3T,3X\}$- Ricci vacuum equation $R^m_q=0$. Being
originally separable in the two orthonormal 3D-time and 3D-space subspaces, those two 3D-local cylindrical geometries are made then to link with each other in a 6D bi-cylindrical formalism by the following means:\\
i/ A more generalized functional cylindrical parameters $\psi(t_i,x_j)$ and $\varphi(t_i,x_j)$ are
extended in both time-like and space-like subspaces (where $\{i,j\}$ are summation indices of curved coordinates). Being chosen in accordance with the traditional quantum physics in a bi-cylindrical
geometry in the symmetrical curved time-space: $\{3T,3X\} \equiv \{\psi,\varphi,t_k \mid\psi,\varphi,x_l\}$, those functional cylindrical parameters are
proposed to be separable into not only time-like $\{\psi_T,\varphi_T\}$ and space-like $\{\psi_X,\varphi_X\}$ constituents, but also linear-longitudinal $\{\psi_{Li},\varphi_{Li}\}$ and curved transverse rotational $\{\psi_{Ro},\varphi_{Ro}\}$ ones, namely:
\begin{eqnarray}
&&\psi(t_i,x_j)\equiv \psi_T(t_0,t_k)\times \psi_X(x_n,x_l)\equiv \psi_{Li}(t_k,x_l)\times \psi_{Ro}(t_0,x_n);\nonumber \\
&&\varphi(t_i,x_j)\equiv \varphi_T-\varphi_X \equiv \varphi_{Li}+\varphi_{Ro}=(\Omega_0 t_0+\Omega_3 t_3)-\nonumber \\
&&-(k_n x_n+k_3 x_3)\equiv \Omega t-k z\equiv \Omega_i t_i-k_jx_j.
\label{eq6}
\end{eqnarray}
Due to separability of the bi-cylindrical functional parameters  $\psi(t_i,x_j)$ and $\varphi(t_i,x_j)$, the 3D-local geodesic conditions of separate cylindrical curvatures will be conserved. The latter is important for consideration of the microscopic cosmological model in each of the 3D-subspaces.   \\
ii/ A Lorentz-like condition is proposed for compensation of any excessive longitudinal fluctuations between time-like and space-like translational
displacements in the original harmonic plane wave equation $(\ref{eq3})$:
\begin{equation}
(\Omega_3^2-k_3^2)\psi\equiv (\Omega_k^2-k_l^2)\psi_0(t_k,x_l)=0,
\label{eq15}
\end{equation}
Implying that $t_3,x_3$ are selected axes of  cylindrical geometry embedded in the flat $\{3T,3X\}$ time-space, i.e. $t_3\in \{t_k\}$ and $x_3\in \{x_l\}$, it
leads to a similar harmonic plane wave equation:
 \begin{equation}
 \frac{\partial^2 \psi}{\partial t_k^2}=\frac{\partial^2 \psi }{\partial x_l^2}.
 \label{eq3a}
  \end{equation}
For conservation of Lorentz invariance, it will be clear that Equation $(\ref{eq3a})$ should be conserved in accordance
with the conservation of linear translation (CLT)  in any lower-dimensional space-time embedded in the 6D-lightcone $(\ref{eq2})$.
In the simplest case of a cylinder with a fixed radius or the latter is adiabatically varying, it is accepted that the absolute value
$\mid\psi(t_k,x_l)\mid = \mid\psi_{Li} \mid= \mid\psi_0(t_k,x_l)\mid = \psi_0=$ constant, where $\psi_0$ is the amplitude of original harmonic plane waves $(\ref{eq3})$.
As a result, $\psi(t_k,x_l)\equiv \psi_0(t_k,x_l)$.    \\
The possibility of separable conservation of the 3D-local geodesic conditions in each of the 3D-orthonormal subspaces allows us to adjust their scale independently. Then assuming that due to interaction of a Higgs-like potential the time-space symmetry is spontaneously broken, leading to the formation of energy-momentum, in particular, to formation of particle masses. As a result, the bi-cylindrical geometry $(\ref{eq5})$ becomes asymmetrical, as shown in Ref.\cite{Vo3}:
\begin{equation}
d\Sigma^2=(ds_0^2+ds_{ev}^2)-(d\sigma_{ev}^2+d\sigma_L^2)=dt^2-dz^2,
\label{eq5a}
\end{equation}
where: $dt^2=d\psi(t_0,t_3)^2+\psi(t_0,t_3)^2d\varphi(t_0,t_3)^2+dt_3^2$ \\
and: $dz^2=d\psi(x_n,x_3)^2+\psi(x_n,x_3)^2d\varphi(x_n,x_3)^2+dx_3^2.$   \\
For generalized functional cylindrical parameters $\psi(t_i,x_j)$ and $\varphi(t_i,x_j)$ in $(\ref{eq5a})$ only the corresponding effective variables in each of the time-like or space-like
subspaces are written explicitly. The time-like and space-like intervals in Geometry $(\ref{eq5a})$ separate into odd terms ($ds_0$ and $d\sigma_L$) and  even terms ($ds_{ev}$ and $d\sigma_{ev}$), which means that the corresponding cylindrical accelerations cannot flip (as an odd-term) or can flip forward and backward (as an even-term) in relation to the cylindrical axis. They are made by a semi-phenomenological modeling to meet the realistic physical interactions of
different intensities (Higgs-like, electro-weak and CPV-ones), which is set up since a moment of spontaneous breaking of time-space symmetry. \\
For the geodesic solutions, it is assumed that the Hubble law of the cosmological expansion is applied to the bi-cylindrical model of microscopic space-time: $\frac{\partial \psi}{\partial y} =v_y=H_y\psi$. Therefore:
$\left[\frac{\partial y}{\partial \psi}\right]=\frac{1}{H_y \psi}$, where $v_y$, the expansion rate is proportional to  the "microscopic scale factor" $\psi(y)$ and $H_y$  is a "microscopic Hubble constant"; where for briefing: $y\equiv\{t,z\}\equiv\{t_0,t_3,x_n,x_3\}\in \{t_i,x_j\}$. As $\{i,j\}$ are summation indices of curved coordinates then $\{t_i\}$ and $\{x_j\}$ are explicitly embedded in 3D-time or in 3D-space, correspondingly. Among original independent time-like and space-like solutions, each
corresponding time-space pair can be combined by the above-proposed bi-cylindrical formalism, which looks like installation of an engine (a time-space
bi-cylindrical solution) from standard structural components (being selected from time-space separated solutions $R^m_m=0,\ m=1-6$), including longitudinal fluctuations along linear translational axes, as well as transverse cylindrical rotations. As a result, two independent solutions of Equation $(\ref{eq7})$ are found in the bi-cylindrical geometrical expression, namely:
\begin{eqnarray}
&&R^3_3(y_i)+R^3_3(y_j)=0; \nonumber \\
&&R^\psi_\psi(y_i)+ R^\psi_\psi(y_j)=0.
\label{eq5f}
\end{eqnarray}
The first of Sub-equations $(\ref{eq5f})$, in combination with the assumption of a Lorentz-like condition leads to the conservation of linear translation (CLT), as expected. In the second independent sub-equation $(\ref{eq5f})$, the acceleration term in 3D-time  is enhanced strongly due to interaction with a Higgs-like potential $V(\phi)$ of which for the major component $V_T$ governed by a time-like "cosmological constant" $\Lambda_T=\left( \frac{\partial \varphi}{\partial t_0}\right)^2$, and relatively, with a much less component $V_X$ acting in 3D-space.
Even the Higgs boson is a scalar in 3D-space, it is not any constraint to assume that a Higgs-like field is simultaneously pseudo-vector in 3D-time subspace, which is able to re-orientate pseudo-spin $\vec{\tau}$ exacting an electromagnetic potential in inducing the ferromagnetism in 3D-space. In a result, the time-like potential $V_T$ creates an absolute polarization toward the future in 3D-time, while the space-like component $V_X$ contributes in polarization of spin $\vec{s}$ in 3D-space. This Higgs-like mechanism turns the higher-dimensional symmetrical bi-geodesic solution $(\ref{eq5f})$ in to an asymmetrical bi-geodesic equation, corresponding to Geometry  $(\ref{eq5a})$:
\begin{eqnarray}
\frac{\partial^2 \psi}{\partial t^2} -\frac{\partial^2 \psi}{\partial {x_j}^2} =\left [\Lambda_T -\left(\frac{\partial \varphi}{\partial x_n}\right)_{even}^2-\Lambda_L \right ]\psi,
\label{eq21}
\end{eqnarray}
where $\Lambda_L\equiv \left(\frac{\partial \varphi}{\partial x_n^L}\right)^2$ is a small P-odd space-like "cosmological constant" due to the global weak interaction in 3D-space. Equation $(\ref{eq21})$ describes the microscopic cosmological geodesic evolution of time-space curvatures by its monotone exponential solution $\psi=\psi_0 e^{\pm \varphi}=\psi_0 e^{\pm (\Omega t+k_j x_j)}$. By the following covariant transformations of time-coordinates: 
\begin{eqnarray}
&& t_i  \leftrightarrow -i.t_i; \ \  \ x_j \leftrightarrow i.x_j \nonumber \\
&& \frac{\partial }{\partial t_i}\rightarrow i\frac{\partial}{\partial t_i}; \  \frac{\partial }{\partial x_j} \rightarrow -i\frac{\partial }{\partial x_j},
\label{eqTrans1}
\end{eqnarray}
one can turn Equation $(\ref{eq21})$ in to a wave-like  representation with $\psi_w\equiv\psi(y\rightarrow iy)\sim e^{i\varphi}=e^{i(\Omega t-k_j x_j)}$. Namely, the corresponding wave-like solution reads:
\begin{eqnarray}
-\frac{\partial^2 \psi}{\partial t^2}+\frac{\partial^2 \psi}{\partial {x_j}^2}=\left [\left(\frac{\partial \varphi}{\partial t_0^+}\right)^2-B_e ( k_n.\mu_e)_{even}^2 -\left(\frac{\partial \varphi}{\partial x_n^L}\right)^2 \right ]\psi,
\label{eq22a}
\end{eqnarray}
where $B_e$ is a calibration factor  and $\mu_e$ is the P-even magnetic dipole moment of charged lepton, correlated with spin vector $\vec s$. Actually, the transformation from the exponential solution to the wave-like one is realized by quantum dynamical operators exacting covariant Transformation $(\ref{eqTrans1})$ in scaling with the Planck constant and the Compton wave length, namely: $\frac{\partial }{\partial t}\rightarrow \hat{E}=i\hbar\frac{\partial}{\partial t}$; $\frac{\partial }{\partial x_j} \rightarrow \hat{p_j}=-i\hbar\frac{\partial }{\partial x_j}$. As a result, a generalized Klein-Gordon-Fock equation is formulated from Representation $(\ref{eq22a})$ as:
\begin{equation}
-\hbar^2\frac{\partial^2 \psi}{\partial t^2}+\hbar^2\frac{\partial^2 \psi}{\partial x_j^2}-m^2\psi=0,
\label{eq23a}
\end{equation}
where the square mass term is:   $m^2=m_0^2-m_s^2$. There beside the rest mass $m_0$, the P-even contribution of $m_s$ is linked with an external curvature due to spinning in 3D-space. At variance with the conventional Klein-Gordon-Fock equation, which is often considered as an equation of motion of particle with spin zero, Equation $(\ref{eq23a})$ serves as a description of spinning particles and due to cylindrical specification, it is most convenient for fermions. In particular, it is reminiscent of the squared Dirac equation of a polarized electron with a given spin projection. Indeed, as the KGF equation is a more universal by its explicit Lorentz covariant formulation. Applying factorization based on Dirac matrices, the second order differential equation $(\ref{eq23a})$ is linearized  in to a Dirac equation for a separated spinor component with its fixed spin projection ($s_n=+1/2$ along the momentum or $s_n=-1/2$ against the momentum).
 \\
When the spinning is not able observed, the curved geometry $(\ref{eq5a})$ turned to an almost exacting formula of special relativity:
\begin{equation}
d\Sigma^2 \Rightarrow ds_0^2=dt_i^2-dx_l^2=dt^2-dx_l^2,
\label{eq5b}
\end{equation}
In accordance with Geometry $(\ref{eq5b})$ the generalized KGF equation $(\ref{eq23a})$ turns to the conventional KGF equation for spin-zero particle:
\begin{equation}
-\hbar^2\frac{\partial^2 \psi}{\partial t^2}+\hbar^2\frac{\partial^2 \psi}{\partial x_l^2}-m_0^2\psi=0,
\label{eq23b}
\end{equation}
\section{A representation of the higher-dimensional Lagrangian}
In Quantum field theories (QFT) the factorized Dirac equation of a genetic massless electron in interaction with the Higgs potential is able to be derived from the following Lagrangian density:
\begin{equation}
L_e=i.\overline{\psi} \left(\gamma^{\mu} \partial_{\mu}-i.f_e \phi \right)\psi-V(\phi),
\label{eqA1.1}
\end{equation}
where $V(\phi)=\eta^2\phi^2+\lambda\phi^4$ is a symmetrical potential; $\lambda < 0$ is the coupling constant of Higgs self-interaction; the constant $\eta$ defines the physical Higgs global vacuum expectation  $\phi_0=\eta (-2 \lambda)^{-{1/2}}$ and $f_e$ is the coupling constant of electron-Higgs interaction. Then a spontaneously symmetry-breaking at $\phi=\pm\phi_0$ can induce a nonzero mass to genetic massless electron. \\
Regarding the higher-dimensional curved geometry $(\ref{eq5a})$, a space-time linearization rule is proposed, namely, the curved EDs are factorized by the SU(2) Pauli matrices $\tau_k$ in 3D-time and $\sigma_l$ in 3D-space. In the result, the cylindrical curvatures are getting hidden under the constant spinning factors $\tau_k \equiv \pm |\vec{\tau}| \equiv \pm [\frac{1}{2}\hbar]$ and $\sigma_l \equiv \pm |\vec s| \equiv \pm [\frac{1}{2}\hbar]$, while the curved coordinates are formally linearized as $dt_0 \rightarrow [\tau_3]dt_3$ and $dx_n \rightarrow [\sigma_l] dx_l$. In a spontaneous symmetry breaking, both vectors $\vec{\tau}$ and $\vec{s}$ are fixed, correspondingly, along a given direction in 3D-time or in 3D-space subspaces. For further definition, the plus sign ($+$) is selected. In analogue to the electromagnetic gauge potential, the ED-induced covariant derivatives can be added to the time-space structure as follows: 
\begin{eqnarray}
&& \left[\frac{\partial}{\partial t_0}-i.f_e\phi(t_0)\right] \equiv \tau_3 \frac{\partial}{\partial t_3}-i.f_e \phi([\tau_3]t_3); \nonumber \\
&& \left[\frac{\partial}{\partial x_n}-i.f_s\phi(x_n)\right] \equiv \sigma_l \frac{\partial}{\partial x_l}-i.f_s \phi([\sigma_l]x_l).
\label{eqA1.2a}
\end{eqnarray}
Correspondingly, their covariant derivatives in the higher-dimensional Lagrangian can be added to the extended linear derivatives in interaction  of massless charged leptons with electromagnetic potentials $\{A_0,A_l\}$:
\begin{eqnarray}
&&L_{6D}= i.\overline{\psi} \gamma^0\left[\left(1+\tau_3\right)\frac{\partial}{\partial t_3}-i.e\left(A_0+\frac{f_e}{e} \phi([\tau_3]t_3) \right)\right]\psi + \nonumber \\
&&i.\overline{\psi} \gamma^l\left[\left(1+\sigma_l\right)\frac{\partial}{\partial x_l}-i.e\left(A_l+\frac{f_s}{e} \phi([\sigma_l]x_l) \right)\right]\psi-V(\phi).
\label{eqA1.2b}
\end{eqnarray}
Being a SU(2)-doublet with four scalar components, the Higgs-like field $\phi=\left\{ \frac{0}{(\chi+\phi_0)} \right \}e^{i.\varphi}$ has, however, only two valid components being able to be involved in the spontaneous time-space symmetry breaking, which are differed from each other by the coupling factors $f_e\gg f_s$. In a result, the first component induces the proper mass $f_e\phi([\tau_3]t_3) \rightarrow + f_e\phi_0=+\Omega_0\equiv m_0$, while for the second one due to having the only scalar for all three Goldstone bosons, it induces an intrinsic mass contribution from spinning along one of the three axes, e.g. $dx_3$ in 3D-subspace, $f_s\phi([\sigma_l]x_l)\rightarrow f_s \phi([\sigma_3]x_3)\rightarrow f_s\phi_0=k_n=m_s$. The two other massless Goldstone bosons make spinning simultaneously sterile along the other two orthogonal axes in 3D-spatial subspace. In accordance with the time-space asymmetrical bi-cylindrical geometry $(\ref{eq5a})$, the extended coordinates are transformed into traditional linear coordinates of 4D space-time as follows:
\begin{eqnarray}
&&\gamma^0\left(1+\tau_3\right)\frac{\partial}{\partial t_3} \rightarrow \gamma^0\frac{\partial}{\partial t}; \nonumber \\
&&\gamma^l\left(1+\sigma_l\right)\frac{\partial}{\partial x_l} \rightarrow  \gamma^j\frac{\partial}{\partial x_j},
\label{eqA1.3b}
\end{eqnarray}
where the 4D-indexes in the left side $\mu=\{0,l\}$ relate to the 6D-linear coordinates $\{k,l\}$, while those in the right side after transformation $\mu=\{0,j\}$ relate to the linearly factorized  coordinates  $\{i,j\}$ with hidden cylindrical curvatures. \\
Proposing further that before symmetry breaking, $\phi=0$,  four-component scalar field $\phi(t_k,x_l)$ can serve as the phase shifts in the gauge transformation of four-vector electromagnetic potential $\{A_0,A_l\}$, correspondingly. Namely, it reads:
\begin{eqnarray}
&&\phi([\tau_3]t_3)=\frac{\tau_3}{f_e}\frac{\partial\varphi([\tau_3]t_3)}{\partial{t_3}}\equiv \frac{1}{f_e}\frac{\partial\varphi(t_0)}{\partial{t_0}}=\frac{\Omega_0}{f_e}; \nonumber \\ 
&&\psi \rightarrow \psi'=\psi e^{i\varphi([\tau_3]t_3)}, 
\label{eqA1.3a1}
\end{eqnarray}
and: 
\begin{eqnarray}
&&\phi([\sigma_l]x_l)=\frac{\sigma_l}{f_s}\frac{\partial\varphi([\sigma_l]x_l)}{\partial{x_l}}\equiv \frac{1}{f_s}\frac{\partial\varphi(x_j)}{\partial{x_j}}=\frac{k_j}{f_s}\equiv\frac{k_n}{f_s}; \nonumber \\ 
&&\psi \rightarrow \psi'=\psi e^{i\varphi([\sigma_l]x_l)}.
\label{eqA1.3a2}
\end{eqnarray}
Now, instead of Lagrangian $(\ref{eqA1.1})$ in accordance with the transformation $(\ref{eqA1.3b})$ from 6D time-space symmetrical geometry to the traditional 4D space-time with the linearly factorized coordinates $\{t_0,x_j\}$. Moreover, without electromagnetic interactions, $A_{\mu}\equiv \{A_0,A_l\}=0$, Lagrangian density  $L_{6D}$ is reduced to 4D-Lagrangian density of a freely moving electron:
\begin{eqnarray}
L_{free}= i.\overline{\psi} \gamma^0 \left( \frac{\partial}{\partial t}-i.\Omega_0 \right)\psi + i.\overline{\psi} \gamma^j \left( \frac{\partial}{\partial x_j}-ik_j \right)\psi-V(\phi_0).
\label{eqA1.2}
\end{eqnarray}
Applying the least action principle one derives finally a generalized Dirac equation as an Euler-Lagrange solution of a freely moving electron with nonzero mass terms $m_0=\Omega_0$ and $ m_j=|k_j|=\{0,0,|k_n|\} \equiv m_s$:
\begin{equation}
i.\gamma^{\mu}(\partial_{\mu}+i.m_{\mu})\psi =i.\gamma^{\mu}\partial_{\mu}\psi-(\gamma^0.\Omega_0+\gamma^j.k_j)\psi= 0.
\label{eqA1.4}
\end{equation}
Multiplying the conjugate operators $-i.\gamma^{\mu}(\partial_{\mu}-i.m_{\mu})$ from the left side of ($\ref{eqA1.4}$), one obtains the squared Dirac equation for each component of the four-spinor as for an independent scalar field:   
\begin{eqnarray}
\gamma^{\mu}(\partial_{\mu}-i.m_{\mu})\gamma^{\mu}(\partial_{\mu}+i.m_{\mu})\psi=&&(\gamma^{\mu}\partial_{\mu})^2\psi+(\gamma^{\mu}m_{\mu})^2\psi= \nonumber \\
&&\partial_0^2\psi-\partial_j^2\psi+(m_0^2-m_s^2)\psi=0.
\label{eqA1.4a}
\end{eqnarray}
Finally, one obtains a generalized KGF equation $(\ref{eq23a})$ (with $\hbar=c=1$) for a spinning electron:
\begin{equation}
-\frac{\partial^2\psi}{\partial t^2}+ \frac{\partial^2\psi}{\partial x_j^2}-(m_0^2-m_s^2)\psi=0.
\label{eqA1.4b}
\end{equation}
Even time axis $dt$ is originally curved, its cylindrical curvature is internal for us as the 4D-observers, i.e. imitating a zero-curvature in our 4D observation. In the internally flat 4D-Minkowski geometry, the longitudinal time-axis $dt$ is synchronized with electron evolution and all the human observations, which base on the same electron evolutional level. It is assumed that all other leptons heavier electron can follow the evolutional time-axis of the electron by projecting the foot-prints of their more complicated curved time-evolution on the physical time axis $dt$. 
Based on QFT formalism, for synchronization with time-evolution of the electron, their quantum wave functions $\psi_{SM}$ as the solutions from SM in the flat 4D-Minkowski space-time may be treated under a gauge-like transformation to get in an internal cylindrical evolution, described by a wave function $\psi$ in the curved time-evolution similar to a solution of the generalized equation $(\ref{eq23a})$, namely:
\begin{equation}
\psi_{SM}\Rightarrow \psi=\psi_{SM}.e^{i\varphi_0}\equiv \psi_{SM}.e^{i\Omega_0.t_0},
\label{eq23c}
\end{equation}
In general, a gauge-transformation ensures the invariance of a Lagrangian in Standard model (SM) of quantum field theories.   Extending the gauge transformation of electro-magnetic potential $A_{\mu}\equiv\{V_0,\vec{A}\}\rightarrow A^{'}_{\mu}=A_{\mu}+\frac{1}{e}\frac{\partial {\varphi(y)}}{\partial y}$ by shifting its time-like component $A_0\rightarrow A_0+\frac{1}{e}\Omega_0$, being equivalent to $(\ref{eq23c})$, that transformation is able to involve all charged leptons in the same time-like cylindrical circulation as one of the electron in according to the higher-dimensional dynamics. As a result, being originally embedded in the more generalized curved geometry $(\ref{eq5a})$, Geometry $(\ref{eq5b})$ finally turns exacting the flat 4D-Minkowski geometry of Special relativity and serves the space-time scene for accommodation of the SM which gets the description of heavier elementary particles in a qualitative consistency with traditional quantum field theories. However, the derived quantum mechanical equation $(\ref{eq23a})$ describes only freely moving charged leptons without accounting their interactions, therefore, it will be a subject of further research to meet a quantitative consistency of higher-dimensional geometrical dynamics with SM physics.
\section{Microscopic cosmological model for calculation of charged lepton mass ratios}
The consistency between the above-mentioned time-space symmetry-based higher-dimensional geometrical dynamics with quantum theories serves a basis for a research on the origin of mass-hierarchy of charged leptons. In duality to the wave-like equation $(\ref{eq23a})$, the geodesic equation $(\ref{eq21})$ in a homogeneous condition (i.e. without the translational terms) leads to an independent geodesic condition separated locally in 3D-time:
\begin{equation}
\frac{\partial^2 \psi}{\partial t_0^2}= \Lambda_T \psi,
\label{eq21T}
\end{equation}
It is equivalent to de Sitter-like solutions $\psi(t_0)=\psi_0(0).e^{\sqrt{\Lambda_T}t_0}$, which may serve for modeling Hubble expansion in the microscopic space-time. 
In the present study with a homogeneous condition, due to experimental uncertainty of the spinning even-term $m_s$ in 3D-space, it is proposed to link the higher-dimensional curvatures lonely with the proper mass $m_0$, as in an approximation.
Up to now the curved time axis has been existing lonely without an explicit embedding in the 3D-time subspace.  
In the so-called microscopic cosmological model the cylindrical curvature as well as all higher order hyper-spherical curvatures up to the 3rd order are implemented. They are described by the same radial parameter $\psi$, therefore, in the assumed Hubble-like expansion the isotropic condition is also taken place. Moreover, it is assumed that the 3D time-like micro-cosmos at the first moment expands exponentially similar to the inflation mechanism. Consequently, based on the original time-space symmetry, it is assumed further that the expansion is getting moderate with the same expansion rates of the relativistic radiation era and the matter dominant era, correspondingly, exacting the Standard cosmological model of the Universe in the macroscopic 3D-space. At variance with the latter, however, the 3D-time micro-cosmos is embedding the hyper-spherical surfaces in a quantum mechanical scale, which re-normalizes amplitude of the radial parameter $\psi$ of time-like hyper-surfaces by the Planck constant and the Compton wave length. In the meantime, the time-like rotational parameter $\varphi$, or its equivalent proper time $t_0$, reveals as an observable macroscopic time variable. Another different feature of microscopic cosmological model is the emphasized role of the first order cylindrical curvature, which is mandatory for all lepton generations as a basic evolution in a coherent relation with the widespread time evolution of electrons.
 Under those concepts, the microscopic cosmological model is proposed to solve the mass hierarchy problem of charged leptons, in which the masses of heavier leptons are determined by observation from the basic macroscopic evolutional time-axis.\\
In Ref.\cite{Vo3} it was proposed that all charged leptons follow the same cylindrical evolution, characterized by polar parameter $S_1(\varphi^+)$, where $\varphi^+=\varphi({t_0}^+)$ is azimuth in the plane $\{t_1,t_2\}$ about the axis $t_3$ and the sign $+$ means an evolution toward the future. The fact that all charged leptons conserve their evolutional direction ensures the causality as a primary important principle of physical reality. This serves also a highly positive argument for justification of the fact, why the electron and its heavier charged triplet partners are look so much like each other, except their absolute masses.
Concerning the Pauli principle, it is mentioned that even though all identical particles, i.g. electrons, are synchronizingly evolving in a coherent time-like microscopic state, similar to a superconducting condensate, they are still in different space-like microscopic positions. 
The mass hierarchy is solved by introduction of higher orders of time-like curvatures, namely, the second order of muon and the third order of tauon. The number of lepton generations is determined by the maximal dimension of 3D-time subspace, namely, fixed with the number "3". Technically the higher hyper-spherical surfaces are described by the nautical angles $\{\varphi^+,\theta_T,\gamma_T\}$, where $\theta_T$ is a zenith in the plane $\{t_1,t_3\}$ and $\gamma_T$ is another zenith in the orthogonal plane $\{t_2,t_3\}$.
Even though all three kinds of leptons have the same time-like radii, their higher-dimensional quantum wave functions are different. It is because they are constructed in variable separable formulas, such as $\psi_{n>1}=\psi_1(\varphi^+)\Theta(\theta,\gamma)$ and $|\Theta(\theta,\gamma)|^2=1$; therefore, the phase variations of heavier leptons ($n>1$), determined by $\Theta(\theta,\gamma)$, are different than one of electrons in $\psi_1(\varphi^+)$. Moreover, even though the time-like radii are proposed identical, the space-like classical radii of three kinds of charged leptons are different, namely, inversely proportional to their corresponding proper mass as shown in Ref.\cite{Vo3}. \\
In the first order of approximation, charged lepton masses are calculated by quite simple formulas:
\begin{eqnarray}
m_n=\rho_n.V_n(\Phi)=\frac{\rho_1}{|\psi|^{n-1}}V_1S_{n-1}=W_1\rho_{n-1}S_{n-1},
\label{eq24}
\end{eqnarray}
where $\rho_n\sim C_n$ is the energy density proportional to the maximal curvature of a $n-$hyperspherical surface $S_n$, then the curvature is inversely proportional to the cylindrical radius of n-power as:  $C_n\sim |\psi|^{-n}$; $V_n$ is the corresponding co-moving volume. The latter is determined by consideration that the higher-order additional curvatures are external to 4D-observers, because the latter are involved only in the lowest curvature level of electrons, then they can recognize the higher additional hyper-spherical surfaces only as the foot-prints projected on the basic cylindrical evolutional axis $dt$. \\ In detail, the energy density of an electron depends on its internal curvature as: $\rho_1=\epsilon_0/\psi$; where $\epsilon_0$ is introduced as a universal lepton energy factor. The co-moving volume of electron is $V_1(\varphi^+)=\Phi=\psi T$, where $T$ is the time-like Lagrange radius, $\Phi$ is the time-like microscopic Hubble radius and $\psi$ serves as the time-like scale factor in the microscopic cosmological model. Therefore, the mass of the electron is determined as:
\begin{equation}
m_1=\rho_1V_1=\rho_1\Phi=\epsilon_0.T=\epsilon_0 W_1.
\label{eq22}
\end{equation}
The value $W_1$ is the time-like Lagrange "volume" of the electron. \\
For the simplest additional configurations $S_1(\theta_T)$ and $S_1(\gamma_T)$ the lepton mass of a 2D time-like curved muon is: 
\begin{eqnarray}
m_2&&=W_1 \rho_1[S_1(\theta_T)+S_1(\gamma_T)]= W_1.\rho_1.[2.S_1]\equiv \nonumber  \\
&&\equiv \rho_2.[2.V_2]=\epsilon_04\pi.T^2 =\epsilon_0 W_2,
\label{eq25}
\end{eqnarray}
where the factor of $2$ of $V_2$ implies that, instead of the general formula $(\ref{eq24})$, the co-moving volume of muon through Equation $(\ref{eq25})$ of $m_2$ consists of two identical hyper-spherical terms. \\
For the simplest additional $S_2(\theta_T,\gamma_T)$ configuration the lepton mass of a 3D time-like curved tauon is:
\begin{equation}
m_3=W_1 \rho_2 S_2(\theta_T,\gamma_T)=\epsilon_04\pi.T^3=\epsilon_0 W_3,
\label{eq26}
\end{equation}
where in $(\ref{eq25})$ or $(\ref{eq26})$ $W_n$ is dimensionless Lagrange volume. There are in the mass formulas  two free parameters: the time-like Lagrange radius $T$ and the lepton energy factor $\epsilon_0$ which are fixed by experimental masses of two from three charged leptons and then used for predicting the mass of the third one. In particular, applying Equations $(\ref{eq22})$ and $(\ref{eq25})$  with the masses of electron and muon for calibration, the two free parameters are found as: $\epsilon_0=31.05603074$ keV and $T=16.45409713$. Then Equation $(\ref{eq26})$ predicts tauon absolute mass (in MeV) in a mass hierarchy, as follows:
\begin{eqnarray}
m_1:m_2:m_3=0.5109989461:105.6583745:1738.51.
\label{eq27}
\end{eqnarray}
While the experimental data of charged lepton masses from the updated PDG \cite{Patr1} give:
\begin{eqnarray}
m_e:m_\mu:m_\tau= 0.5109989461(31):105.6583745(24):1776.86(12).
\label{eq28}
\end{eqnarray}
The predicted tauon mass is consistent with experimental data within $2.16\%$ of standard deviation. \\
In \cite{Vo10} a fine-tuned calculation has been implemented by adding higher-order corrections of the curvatures to the major terms led to a better consistency almost for 85 times. In particular, the mass of tauon reaches a quantity $m_{\tau}(theor)=1776.40$ (MeV). \\
In accordance with the fine-tuning calculation \cite{Vo10} of the mass ratios of charged leptons, one finds the mixing of lower curvatures to the major ones as the corrections, come from electron to muon mass and from electron and muon to tauon mass. 
The phase shift of gauge transformation in the first order with a phenomenological coupling constant $f_1=f_e$ in $(\ref{eqA1.3a1})$ is applied to induce the proper mass of electron. In the same way, one can add the second and the third orders of the gauge phase shifts with couplings $f_{\mu}$ and $f_{\tau}$ to induce the masses to muon and tauon, while conserving the first order shift based on the gauge invariance. This leads to the following gauge transformer for muon mass:
\begin{eqnarray}
f_e\phi(t_0) \rightarrow f_{\mu}\phi(t_0) &&=f_e\left(1+\frac{f_2}{f_1}\right)\frac{\partial\varphi(t_0)}{\partial{t_0}}\Rightarrow (f_1+f_2)\phi_0\equiv m_{\mu}; \nonumber \\ 
&&\psi \rightarrow \psi'=\psi e^{i. [f_{\mu}\phi(t_0)]t_0}=\psi e^{i \Omega_{\mu}t_0}, 
\label{eqA1.3a1ED}
\end{eqnarray}
and similar for tauon mass: 
\begin{eqnarray}
f_e\phi(t_0) \rightarrow f_{\tau}\phi(t_0) &&=\left(f_e+\frac{1}{2}f_2+f_3\right)\frac{\partial\varphi(t_0)}{\partial{t_0}}\Rightarrow \left(f_1+\frac{1}{2} f_2+f_3\right)\phi_0\equiv m_{\tau}; \nonumber \\ 
&&\psi \rightarrow \psi'=\psi e^{i. [f_{\tau}\phi(t_0)]t_0}=\psi e^{i \Omega_{\tau}t_0}.
\label{eqA1.3a2ED}
\end{eqnarray}
Thus, both factors $f_{\mu}=(f_e+f_2)$ and $f_{\tau}=\left(f_e+\frac{1}{2} f_2+f_3\right)$ contain the electron mass component $m_e\equiv f_e\phi_0 \sim f_1$ proportional to a basic cylindrical curvature in mass creation of all charged leptons. Serving as new physical parameters, the masses of higher-order curvatures of muon, $m_{\mu}\equiv \Omega_{\mu}$, and tauon, $m_{\tau}\equiv \Omega_{\tau}$, then phenomenologically replace the electron mass in quantum mechanical equations similar to the KGF equation $(\ref{eq23a})$ of electrons. The gauge transformation $\psi \rightarrow \psi'$ in Equations $(\ref{eqA1.3a1})$, $(\ref{eqA1.3a1ED})$ and $(\ref{eqA1.3a2ED})$ implies that the effective wave functions of the corresponding KGF equations are different from each other for three charged lepton generations, as required in quantum theories. 
\section{Discussions}
The mass ratios of fermion elementary particles are calculated by correlation of  flavor mixing angles of quarks in the frame of SM or of neutrinos in an extended SM, which achieved qualitative mass ratios of quarks \cite{{HaHa1},{BrSi1},{FriXi1},{FiHu1},{XiYaZh1}} and some extensions to charged leptons and neutrinos \cite{{XiYaZh1},{Koi2},{Koi3},{HoSa1},{FriXi2},{FriXi3},{FriXi4},{Ishi1},{Alta1},{Har1},{Xin1},{Ky1}}. In particular, the observed mismatch of the mass hierarchy between the up-type and down-type quarks \cite{{Patr1},{XiZhZh1},{XiZhZh2}}, namely, $m_u/m_c\sim m_c/m_t \sim \lambda^4$ while $m_d/m_s \sim m_s/m_b \sim \lambda^2$, where the parameter $\lambda=0.2$, would be determined by parametrization of their mixing matrices with different mixing angles  \cite { FriXi1} or explained by a dynamical difference of their charges \cite { FiHu1}. The assumptions of those correlations are possibly applied to determine mass ratios of the charged lepton sector \cite {{Koi2},{Koi3},{HoSa1},{FriXi2},{FriXi3}}, which qualitatively meet the experiments. In particular, for the assumption of quark-to-lepton correspondences \cite {{Koi3},{FriXi2}}, the charged leptons would be studied in correlation with quarks, that leads to their relatively strong mass hierarchy, roughly of the orders as of the quark sectors, namely, $m_e/m_{\mu}\sim 3.\lambda^4$ and $m_{\mu}/m_{\tau}\sim 4/3.\lambda^2$. \\
Regarding the Koide empirical formula \cite {Ko1}:
\begin{equation}
m_e+m_{\mu}+m_{\tau}=\frac{2}{3}(\sqrt{m_e}+\sqrt{m_{\mu}}+\sqrt{m_{\tau}})^2,
\label{eqKoide}
\end{equation}
interestingly, the prediction of tauon mass by Formula $(\ref{eqKoide})$, based on electron and muon masses, leads to the quantity $m_{\tau}(Koide)=1776.97$ MeV which is in excellent agreement within $1\sigma$ with the experimental tauon mass $m_{\tau}(exp)= \\ 1776.86(\pm 0.12)$ MeV. In Ref.\cite{Koi2} Koide suggested the democrate family mixing, which would lead to the mass ratios of charged leptons. Recently, the Koide formula $(\ref{eqKoide})$ is proposed to extend for the neutrino and quark sectors, as a more universal assumption \cite{{Gao1},{LiMa1}}. In particular, Kartavtsev \cite {Kar1} reformulated a Koide-like empirical formula including all six leptons, and another formula including all six up-type and down-type quarks. Those interpretations do not overcome the phenomenological level. 
Some other geometrical interpretation of the Koide formula was proposed in Ref.\cite{Koc1} where mass correlations are expressed through Descartes-like circles or with their corresponding squared curvatures. For overcoming an empirical level, it needs to propose a new theoretical basis to derive the Koide formula from other new higher-order symmetries common for fermion sectors. Recently, the model proposed by Sumino \cite {Sumi1} based on adding the family gauge bosons for interpretation of the Koide formula within EFT could link the charged lepton masses with the squared vacuum expectations of new gauge bosons. This model is waiting for upgrading to higher order symmetry, which is a challenge due to various assumptions made at very high energies of $10^2-10^3$ TeV. As the Sumino model was designed to deal only with charged leptons, Koide and Nishiura \cite{KoNish1} proposed an extension of the family symmetrical group to describe masses and mixing of quarks and neutrinos. The Sumino model with extended symmetries, in the meantime, would offer new physics including prediction of heavy gauge bosons and new kinds of lepton flavor violation decays for experimental verification above the effective energy cut-off of $10^3$ TeV scale of the next accelerator generation. \\
Under the circumstances, our calculation of mass hierarchy of charged leptons based on the proposed microscopic bi-cylindrical model is an alternative following the beyond SM approach. It is found that all charged leptons correctly follow the basic and universal cylindrical evolution, while their masses are dominantly determined by the highest orders of hyper-spherical curvatures beyond the cylindrical geometry \cite{{Vo3},{Vo10}}. The common time-like cylindrical spiral evolution makes the time-arrow oriented one-way directional, namely, toward to the future, which ensures conservation of
the causality principle as a primary requirement in formulation of a new space-time concept. The proposed cylindrical model would be a solution to deal with the puzzle of why except their mass hierarchy, all generations of charged leptons have very similar physical properties, such as: additive charges, spin, magnetic moments, non-nuclear interaction, beta-decay mechanism, etc. \\
The duality of the higher-dimensional general relativity equation  can shed light on the locality of interaction of elementary particles, because the classical geodesic sub-solution describes a point-like particle which is able to contact other particle locally. Simultaneously, the non-locality of the same microscopic substance described by the dual wave-like solution, namely the KGF equation, is in accordance with quantum mechanical behavior in motion, evolution and interactions of the particle. 
The higher-dimensional geodesic description of a point-like elementary particle seems to turn back to the classical mechanical determinism which would violate the quantum statistical behavior. Indeed, the consistency of this microscopic geometrical approach with the statistical feature of quantum theories can be explained by the collective principle of an experimental observation. 
According to the intrinsic time-like cylindrical curvature \cite{Vo3}, a circulation can be quantized in a minimal portion of time evolution in a period $T_S$ as following:
\begin{equation}
\varphi_0=\Omega_0 t_0=2 n \pi,
\label{eqQ1}
\end{equation}
where $n$ is integer and $\varphi=\varphi_{min}=2\pi=\Omega_0 T_S$.
Indeed, a particle detector being always macroscopic object, contains a huge number of electrons $e^-(i)$ which synchronize with each other in the direction of time evolution, but their individual phases may shift from each other within a period $2\pi$. When one or some of those electrons interact with the registered object, namely, an elementary particle, the moment of registration being distributed statistically in accordance with the phase distribution of the ensemble of electrons $e^-(i)$ in the detector-sensor, is determined by the squared average: 
\begin{equation}
dt^2 \equiv <{dt(i)}^2>={dt_0^+}^2+<{dt_3(i)}^2>,
\label{eqdT}
\end{equation}
where the statistical summation is done over all electrons with index $i$, interacting with the registered particle. They are synchronized by the cylindrical evolution $dt_0$, but would be shifted from each other in $dt_3(i)$ within the period $T_s=\varphi_{min}/2\pi$. As far as the physical detector is a macroscopic mechanism, its statistical collective feature as a principal concept can not be avoided.    \\
The synchronization of charged leptons is not enough for avoiding a significant space-time curvature. Regarding the visible baryon matter,  Nuclei consist of quarks, which are confined and have not yet proven to be in synchronization with electrons. However, there are some arguments for nuclei to be consistent with the synchronization. Firstly, charged leptons don't have nuclear force, so that human sense should based on electromagnetic interaction with electron shells of neutral atoms or molecules. Secondly, when nuclei are opened in ionization, they have integer units of electric charge, moreover, they are well observable in electromagnetic interaction with charged leptons. Finally, most of nuclei are coexisting with electrons in an electric charge balance, namely, in formulation of neutral atoms and molecules with certain binding energy. Those facts imply a synchronization between nuclei and electrons, when a human observation is carried out. Indeed, in the present context, one can assume a synchronization between electrons and nuclei in all neutral atoms and molecules,  at least statistically in a period $T_s$, and as they are dominant compositions in the Universe, most of visible matter should be synchronized. This assumption would reduce the evidence of a global space-time curvature to an upper limit of the cosmological constant $\Lambda$. From other perspective, one can not exclude that there are possible different kinds of matters or simply, space-time curvatures, being unsynchronized with electrons, however by the same reason, they are not easily observable. As a result, those space-time excitations would serve new candidates to solve the problem of dark matter or dark energy.  
\\ 
The proposed bi-cylindrical microscopic cosmological model with our recent fine-tuning calculation of tauon mass in \cite {Vo10} approaching to a $3\sigma$-level quantitative prediction of $m_\tau(theor)=1776.40$ MeV is getting in fairly passable consistency with the experimental mass hierarchy of charged leptons. 
\section{Conclusions}
Beside our previous quantum mechanical interpretations, in the present study, there some more details of insight of the higher-dimensional geometrical dynamical approach are analyzed and emphasized, namely: \\
- The specific property of microscopic cylindrical curvatures ensures the conservation of causality and makes a natural consistency between the curved higher-dimensional time-space with the flat 4D-Minkowski space-time. In particular, a synchronization of electrons with a major part of visible matter in a global scale ensures the consistency of the microscopic cosmological model with a low upper-limit of the cosmological constant, leading to a well-observed flatness of the Universe.  \\
-The collective origin of the macroscopic detection mechanism and the universal role of coherent electron ensembles in a physical observation ensure conservation of statistics, as an objective feature of quantum mechanics. \\
- Understanding of the wave-particle duality sheds light on the duality of the locality of interaction of a point-like massless particle in a classical curved geodesic description and the non-locality of the same quantum substance, but getting massive in the flat 4D space-time in a dual wave-like KGF equation. Moreover, the higher-dimensional cylindrical geometrical dynamics can be proven to be in consistency with the QFT Lagrangian formalism of freely moving charged leptons. \\
Those achievements would add more positive arguments or evidences for a qualitative consistency between the proposed higher-dimensional cylindrical dynamical model with the quantum field theories, which implies some advantages of the classical geodesic sub-solution in complementarity to the dual wave-like sub-solution. \\
As here presented Lagrangian formalism of charged lepton quantum fields in the flat 4D-Minkowski space-time has not yet accounted other their dynamical interactions, and number of our conclusions base on hypothetical assumptions, the achievements of this model are still preliminary. However, this general relativity alternative approach, beyond SM, can be considered at least as an effective semi-phenomenological model serving a powerful tool for interpretation of quantum field theories. In particular, this leads to a quantitative solution of the long-standing problem of mass hierarchy of charged lepton generations and offers an explanation of the similarity of their properties.

\section*{Acknowledgment}
The author thanks deeply Nguyen Anh Ky (Institute of Physics, VAST) for useful discussions, T. Hoang Si, V. Nguyen The (VINATOM) and N.B. Nguyen (Thang Long University) for technical assistance.

\end{document}